\documentclass[12pt]{article} 

\usepackage{mathtools}
\usepackage{amsmath}
\usepackage{amsthm}
\usepackage{multirow}
\usepackage[export]{adjustbox}
\usepackage[left=1in, right=1in, top=1in, bottom=1in]{geometry} 
\usepackage{pdflscape}
\usepackage{booktabs}
\usepackage{rotating}
\usepackage{rotfloat}
\usepackage{float}
\usepackage{subfig}
\usepackage{url}
\usepackage{pdfpages}
\usepackage{rotating}
\usepackage{setspace}
\usepackage{csvsimple}
\usepackage{array}
\usepackage{wrapfig}
\usepackage[utf8]{inputenc}
\usepackage{tabularx}
\usepackage{float}

\usepackage{amsthm}                
\usepackage{amssymb}
\usepackage{statmath}

\usepackage{multirow}
\usepackage{ragged2e}
\usepackage{rotating}

\usepackage{caption}

\newtheorem{theorem}{Theorem}
\newtheorem*{newthm*}{\normalfont\scshape Theorem}
\newtheorem{lemma}{Lemma}

\usepackage{amsfonts}
\usepackage{amsmath} 
\usepackage{tikz}
\usetikzlibrary{positioning,shapes.geometric}
\usepackage{comment}
\usepackage{xcolor}
\usepackage{longtable}
\usepackage{setspace}

\usepackage [english]{babel}
\usepackage [autostyle, english = american]{csquotes}
\MakeOuterQuote{"}

\usepackage{mathrsfs}

\usepackage[labelfont=bf, labelsep=period]{caption}

\makeatletter
\newcommand*{\indep}{%
  \mathbin{%
    \mathpalette{\@indep}{}%
  }%
}
\newcommand*{\nindep}{%
  \mathbin{
    \mathpalette{\@indep}{\not}
  }%
}
\newcommand*{\@indep}[2]{%
  \sbox0{$#1\perp\m@th$}
  \sbox2{$#1=$}
  \sbox4{$#1\vcenter{}$}
  \rlap{\copy0}
  \dimen@=\dimexpr\ht2-\ht4-.2pt\relax
  \kern\dimen@
  {#2}%
  \kern\dimen@
  \copy0 
} 
\makeatother

\usepackage{xr-hyper}

\definecolor{forestgreen}{RGB}{34,139,34}
\usepackage{hyperref}
\hypersetup{
    colorlinks=true,
    linkcolor=black,
    filecolor=black, 
    citecolor=forestgreen,      
    urlcolor=blue
}

\usepackage{sectsty}
\sectionfont{\large}

\usepackage{layout}

\usepackage{geometry}

\newcolumntype{C}[1]{>{\centering\arraybackslash}p{#1}}

\usepackage{authblk}

\usepackage{authblk}
\usepackage{tabularray}
\usepackage{colortbl}

\usepackage{afterpage}
\usepackage{lscape}
\usepackage{graphicx}

\usepackage{datetime}

\makeatletter
\newcommand*{\addFileDependency}[1]{
  \typeout{(#1)}
  \@addtofilelist{#1}
  \IfFileExists{#1}{}{\typeout{No file #1.}}
}
\makeatother

\usepackage[backend=biber,style=nature,]{biblatex}
\usepackage{etoolbox}
\patchcmd{\endproof}
  {\popQED}
  {\par\popQED}
  {}
  {}

\theoremstyle{plain}
\newtheorem*{lemma*}{Lemma}

\begin{document}

\title{Constructing external comparator groups via transportability in mean or in effect measure} 

\author[1,2]{Lawson Ung}
\author[1,3]{Guanbo Wang}
\author[4]{Sebastien Haneuse}
\author[1,2]{Sonia Hernandez-Diaz}
\author[1,2,4]{Miguel A. Hern\'an}
\author[1,2,4,5]{Issa J. Dahabreh}

\affil[1]{CAUSALab, Harvard T.H. Chan School of Public Health, Boston, MA}
\affil[2]{Department of Epidemiology, Harvard T.H. Chan School of Public Health, Boston, MA}
\affil[3]{The Dartmouth Institute for Health Policy and Clinical Practice, Dartmouth Geisel School of Medicine, Lebanon, NH}
\affil[4]{Department of Biostatistics, Harvard T.H. Chan School of Public Health, Boston, MA}
\affil[5]{Richard A. and Susan F. Smith Center for Outcomes Research, Beth Israel Deaconess Medical Center, Boston, MA}

\maketitle{}

\noindent \textbf{Address for correspondence:} Dr. Lawson Ung, Department of Epidemiology, Harvard T.H. Chan School of Public Health, Boston, MA 02115; email: \href{mailto:lawson_ung@hsph.harvard.edu}{lawson\_ung@hsph.harvard.edu}; phone: +1 (617) 495‑1000.
\thispagestyle{empty}

\label{firstpage}
\newpage 
\begin{abstract}\linespread{1.3}\selectfont
Learning about causal effects in target populations and their subsets may be facilitated by combining information from multiple sources. One major class of study designs that combine information involves appending an index study with data from an \textit{external comparator}, which may facilitate head-to-head comparisons of treatments initially studied in different populations. We delineate external comparator analyses under two distinct, but related, identification strategies. The first strategy relies on exchangeability (transportability) of potential outcome \textit{means}, which uses information only on the treatments that are to be compared. The second strategy relies on transportability in \textit{effect measure}, requiring additional use of information on a third treatment common to the populations that have been combined. In a time-fixed setting with a point treatment and non-failure time outcome, we examine identification and estimation under a basic setup where information from an index trial is combined with a second, and external to the index trial, data source. We propose estimators for identifying observed data functionals, with a particular focus on semiparametric efficient augmented weighting estimators that incorporate models for the probability of trial participation, the probability of treatment, and conditional outcome means. We derive the asymptotic properties of these augmented weighting estimators -- including robustness to model misspecification and slower rates of convergence for some nuisance function models -- and use simulation to compare their finite sample performance to estimators based only on outcome modeling or weighting. Last, we provide a practical demonstration of the proposed methods by combining the ACCEPT and PHOENIX 1 randomized trials to evaluate the effect of various biologic agents on plaque psoriasis, a chronic inflammatory disorder. 
\end{abstract} 

\noindent \textbf{Key words:} external comparators; combining information; mean transportability; transportability of effect measures; robust estimation; semiparametric efficiency

\maketitle


%

\newpage
\section{INTRODUCTION}
\label{s:intro}
Studies that combine information from different sources are often motivated by the desire to compare treatments initially evaluated in different populations. Such analyses, sometimes described as \textit{external comparator} studies, may be pursued under two distinct, but related, identification strategies that rely on different exchangeability conditions. A first approach to identification assumes transportability of potential outcome \textit{means}, using information only on the treatments that are to be compared. The second makes the generally weaker assumption of transportability in \textit{effect measure}, requiring additional use of information on a third treatment common to the populations at hand \cite{ung2026combining}. 

More concretely, consider the landmark ACCEPT (NCT00454584 \cite{griffiths2010comparison}) and PHOENIX 1 (NCT00267969 \cite{leonardi2008efficacy}) randomized trials that evaluated the effect of biologic therapies for plaque psoriasis, a chronic autoimmune disorder. The ACCEPT trial compared the tumor necrosis factor inhibitor etanercept against the interleukin 12/23 inhibitor ustekinumab, while PHOENIX compared ustekinumab to placebo. In this setting, one could compare etanercept to placebo under transportability of outcome means between the trial populations, and use information only on the treatment groups of interest. By contrast, invoking the assumption of transportability in effect measure between trial populations requires additional use of the ustekinumab arm common to both trials. While the notion of combining information is not new in clinical practice \cite{guyatt1993users, dans1998users}, the early literature on combining trials \cite{gehan1974non, pocock1976combination}, and evidence synthesis methods such as network meta-analyses \cite{bucher1997results, lumley2002network, caldwell2005simultaneous, ioannidis2009integration}, addressing such comparisons in explicitly causal terms using potential outcomes \cite{rubin1974, robins1986, splawaneyman1990} using individual-level health data is a more recent development \cite{shook2024fusing, zivich2024hiv, ung2026combining}. 

In this paper, we propose identification strategies and estimation procedures when constructing external comparator groups via transportability in mean or in effect measure. Working in the setting of a time-fixed treatment and non-failure-time outcome, we propose a target population and sampling model, causal model, identifiability conditions, and identification approaches to express relevant causal estimands as functions of the observed data. We propose estimators, focusing on semiparametric efficient augmented weighting estimators that have appealing asymptotic properties -- including consistency for their target parameters even if some combination of its nuisance models are misspecified or have slower rates of convergence \cite{robins1994estimation, rotnitzky1998semiparametric, bang2005} -- and use simulation to compare their finite sample performance against other regular and asymptotically linear estimators based only on outcome modeling or weighting. Last, we combine the aforementioned ACCEPT and PHOENIX trials to provide a practical demonstration of the proposed methods to evaluate the effect of biologic therapies in plaque psoriasis.

\section{SAMPLING AND DATA STRUCTURE}

\subsection{Sampling} Using a superpopulation framework  \cite{robins1987foundations, robins1988confidence, robins2002covariance}, we assume a \textit{non-nested} sampling scheme in which data from an \textit{index} study (e.g., a randomized trial) are combined with external data collected separately in another experimental or observational setting. The index and external data are considered simple random samples from strata of some infinite target population defined by participation in the index study \cite{dahabreh2020transportingStatMed, dahabreh2021studydesigns}. We refer to these strata as the index and external populations, respectively. Under non-nested sampling, potential outcome means and average treatment effects are usually identifiable only within subsets of the target population defined by index trial participation, since the sampling probability is generally not known to investigators. For a diagrammatic illustration, see \cite{dahabreh2021studydesigns}.

\subsection{Data structure} Denote $X$ as a vector of baseline covariates; $S$ for trial participation ($S = 1$ for the index trial; $S=0$ for the external data); $A$ for discrete treatment assignment levels; and $Y$ for a binary, count, or continuous outcome. We use $\mathcal A_{S=s}$ to denote the treatments in the index and external populations. A composite dataset (Table \ref{tab:datastructure}) is formed by combining data from the index trial with the external data \cite{dahabreh2020transportingStatMed}, comprising independent, and within levels of $S$, identically distributed realizations of the random tuple $O_i \equiv (X_i, S_i, A_i, Y_i)$ for $i=1,2,3,\dots,n$, where $n$ is the total number of individuals within the index and external data (that is, $n_{S=1}+n_{S=0}=n$). The distributions underlying the index trial participants and non-participants will vary in most applications, even if the eligibility criteria defining the index and external data are the same. For example, there may be variation between the marginal densities of baseline covariates, $f(x| S=1)$ and $f(x| S=0)$; conditional densities for treatment given covariates $f(a| x,S=1)$ and $f(a| x,S=0)$; and conditional densities of the outcome given covariates and treatment, $f(y| x,a,S=1)$ and $f(y| x,a,S=0)$. We consider a setting where the index trial with treatments $\mathcal A_{S=1} = \{0,1\}$ is combined with an external trial with treatments $\mathcal A_{S=0} = \{0,2\}$, but our results extend to settings involving observational data and multilevel treatments. Probabilities and expectations are defined with respect to the distribution induced by combining populations, unless stated otherwise.

\section{IDENTIFICATION}

\subsection{Causal model}

We use potential (counterfactual) outcomes \cite{rubin1974, robins1986, splawaneyman1990, robins2000d} to specify the causal model. We say that consistency holds for the following counterfactual random variables: $A^{s}$ is the counterfactual treatment under intervention to set trial participation $S$ to $s$, where $S \in \{0,1\}$ and $\mathcal{A}_{S=1} \cup \mathcal{A}_{S=0} =\{0, 1, 2\}$; and $Y^{s,a}$ is the counterfactual outcome under joint intervention to set trial participation $S$ to $s$ and treatment $A$ to $a$. Implicit in the claim of consistency is that the definitions of these counterfactual random variables are aligned across data sources, or at least subject to version irrelevance with respect to the counterfactual outcome \cite{vanderWeele2009}. 

The causal model we entertain differs from most causal models described (or implied) in other work on combining information, in that we view index trial participation $S$ as encompassing a set of actions implemented within the trial, \textit{beyond} the act of treatment assignment itself \cite{dahabreh2019identification, ung2025generalizing}. Such actions may include measures to improve adherence, regular contact with the research team, and Hawthorne effects \cite{landsberger1958}. We make this explicit within the causal model, even though we work under the simplifying assumption that the effect of trial participation $S$ on outcome $Y$ is mediated exclusively by treatment $A$, such that the counterfactual outcome under joint intervention on $S$ and $A$ is equivalent to the counterfactual outcome under intervention to set treatment $A$ to $a$, that is $Y^{s,a}=Y^a$. Such an assumption can be relaxed under alternative assumptions that we do not address here in this report \cite{ung2025generalizing, ung2026combining}. Furthermore, while we do not present causal graphs, an extensive literature now exists on how such methods can be used in practical settings \cite{robins1986, greenland1999causal, pearl2009causality, richardson2013primer, bareinboim2016causalfusion}. Finally, we do not consider imperfect adherence, censoring, or competing events; extensions to address such complexities \cite{dahabreh2019identification, dahabreh2022adherence} are forthcoming but not essential to arguments herein.

\subsection{Causal estimands} \label{s:estimands}

Under a non-nested sampling scheme, we wish to identify potential outcome means under intervention to set treatment $A$ to $a=1$ and $a=2$ within the index population, that is $\E[Y^{a=1}| S=1]$ and $\E[Y^{a=2}| S=1]$. Furthermore, we are interested in identifying the average treatment effect on the causal mean difference scale, captured in $\E[Y^{a=1}-Y^{a=2}| S = 1]$. For succinctness, we do not present results on the causal mean ratio scale; these results follow immediately once potential outcome means are identified.

\subsection{Identifiability conditions} \label{s:assumptions}

In this section, we state the identifiability assumptions required to express the causal estimands of interest as functions of the observed data. These assumptions are commonly invoked in causal studies that use data from single \cite{rubin1974, robins1986} and multiple sources \cite{bareinboim2016causalfusion, dahabreh2023efficient}. First, as articulated in our description of the causal model, we assume as preliminary the \textit{absence of trial or study engagement effects} \cite{dahabreh2022adherence, dahabreh2019identification}: for each individual $i$, each $s \in \{0,1\}$, and each $a \in \mathcal A_{S=1} \cup \mathcal A_{S=0}$,  if $A_i = a$, then $Y_i^{s,a} = Y_i^a.$ The remaining identifiability assumptions are applied repeatedly throughout.\\

\noindent \textbf{(A1)} \textit {Consistency of potential outcomes.} For individual $i$ and each $a \in \mathcal A_{S=1} \cup \mathcal A_{S=0}$, if $A_i = a$, then $Y_i^a = Y_i$.\\

\noindent \textbf{(A2)} \textit {Conditional mean exchangeability over treatment $A$ in the index trial population.} For each $x$ with positive density $f(x,S = 1) > 0$ and each $a \in \mathcal A_{S=1}$, $$\E[Y^a | X = x , S = 1 ] = \E[Y^a |  X = x, S = 1 , A = a].$$

\noindent \textbf{(A3)} \textit {Positivity of treatment in the index trial population.} For each $x$ with positive density $f(x, S=1) > 0$ and each $a \in \mathcal A_{S=1}$, $\Pr[A = a |  X = x, S = 1] > 0$.\\

\noindent \textbf{(A4)} \textit {Conditional mean exchangeability (transportability) over trial participation $S$.} For each $x$ with positive density $f(x, S=1) > 0$ and for some $a \in \mathcal A_{S=1} \cup \mathcal{A}_{S=0}$, $$\E[Y^a| X = x, S = 1] = \E[Y^a| X = x, S = 0].$$ This exchangeability condition is a transportability assumption in that it allows information from one population to be used to learn about others \cite{bareinboim2012transportability, dahabreh2021studydesigns}.  A4 also implies transportability of difference effect measures, described below.\\

\noindent \textbf{(A4')} \textit {Conditional exchangeability (transportability) of difference effect measures over trial participation $S$.} For each $x$ with positive density $f(x,S=1) > 0 $ and $a,a' \in \mathcal{A}_{S=1}\cup\mathcal{A}_{S=0}$, $$\E[Y^{a} - Y^{a'} |  X = x , S = 1] = \E[Y^{a} - Y^{a'} |  X = x, S = 0].$$

\noindent \textbf{(A5)} \textit{Positivity for the external population.} For each $x$ with positive density $f(x, S = 1) > 0$, $\Pr[ S = 0 |  X = x ] > 0$.\\

\noindent \textbf{(A6)} \textit {Conditional mean exchangeability over treatment $A$ in the external population.} For each $x$ with positive density $f(x,S = 0) > 0$ and $a \in \mathcal A_{S=0}$, $$\E[Y^a | X = x , S = 0 ] = \E[Y^a |  X = x, S = 0 , A = a].$$

\noindent\textbf{(A7)} \textit {Positivity of treatment in the external population.} For each $x$ with positive density $f(x, S=0) > 0$ and $a \in \mathcal A_{S=0}$, $\Pr[A = a |  X = x, S = 0] > 0$.

\subsection{Identification}
\label{s:identification}

We now provide identification results based on the g-formula \cite{robins1986}, drawing on results that have been presented in our previous work \cite{ung2026combining}. All proofs and weighting re-expressions are in Section 3 of the Supplement. We index some statistical estimands with dual subscripts that correspond to realizations of trial participation $S$ and treatment $A$, for example when defining the functional $\gamma_{s,a}$ using a standard Lebesgue-Stieltjes integral $$\gamma_{s,a}\equiv \int \E[Y|X=x,S=s,A=a]dF(x|S=1) \equiv \E \big[ \E [Y|X, S=s, A = a] \big| S = 1 \big],$$
where the conditional outcome mean $\E[Y|X=x,S=s,A=a]$ is integrated with respect to the cumulative distribution function $F(x|S=1)$.

\subsubsection{Potential outcome mean in the index trial}

We first provide a standard result for the potential outcome mean under intervention to set treatment $A$ to $a=1$ in the index population, requiring data only from the index trial.

\begin{lemma}
Suppose conditions A1, A2, and A3 hold for $a=1$. The potential outcome mean under intervention to set treatment $A$ to $a=1$ within the index population, $\E[ Y^{a=1}| S = 1]$, is identified with $\gamma_{1,1} \equiv \E \left[ \E [Y|X, S=1, A = 1] \big| S = 1 \right].$
\end{lemma}

$\gamma_{1,1}$ is used in all functionals for the average treatment effect, $\E[Y^{a=1}-Y^{a=2}|S=1]$, obtained under identification based on either transportability in mean or in effect measure.

\subsubsection{Treatment comparisons under transportability in mean}

We now consider identification of the potential outcome mean under intervention to set treatment $A$ to $a=2$ within the index population. We first present identification results under transportability in potential outcome mean between the index and external populations (A4).

\begin{lemma} 
Suppose conditions A1, A4, A6, and A7 hold for $a=2$, in addition to condition A5. The potential outcome mean under intervention to set treatment $A$ to $a=2$ in the index population, $\E[ Y^{a=2}| S = 1]$, is identified by $\gamma_{0,2} \equiv \E \big[ \E [Y|X, S = 0, A = 2] \big| S = 1 \big].$
\end{lemma} 

Under conditions for Lemmas 1 and 2, the average treatment effect comparing intervention to set treatment $A$ to $a=1$ vs. $a=2$ within the index population is immediate.

\begin{theorem}
If the conditions required for Lemmas 1 and 2 hold, the average treatment effect comparing intervention to set treatment $A$ to $a=1$ versus $a=2$ within
the index population, $\E[ Y^{a=1} - Y^{a=2} | S = 1]$, is identified by $\psi \equiv \gamma_{1,1} - \gamma_{0,2}.$
\end{theorem} 

Readers may note that there is no testable restriction imposed on the law of the observed data to identify $\psi$, precisely because no information was used on $A=0$ from the index or external data.

\subsubsection{Treatment comparisons under transportability in effect measure}

In contrast, external comparisons under transportability in effect measure require additional information about shared treatments between the index and external populations. Here, one replaces the assumption of transportability in mean with transportability in \textit{effect measure} (A4') \cite{vanderweele2009distinction, dahabreh2023efficient, ung2025generalizing}. The assumption that conditional effect measures are exchangeable (transportable) over trial participation $S$ is weaker than that of exchangeable conditional means, in the sense that the latter implies the former, but not necessarily vice versa. In applied generalizability and transportability analyses for example, clinical intuition often suggests that it is more reasonable to assume that risk differences or ratios from clinical trials, rather than absolute risks, can be applied to target populations \cite{glasziou1995evidence, dahabreh2024relative, wang2024relative}. Here, we consider identification under transportability in difference effect measures; results using transportability in relative effect measure are provided in Section 3 of the Supplement. 

First, we provide the identification result for the potential outcome mean under intervention to set treatment $A$ to $a=2$ within the index population. 

\begin{lemma}\label{lemma3}
Suppose conditions A1, A4', A6, and A7 hold for $a=0,2$; conditions A2 and A3 hold for $a=0$; and A5. The potential outcome mean under intervention to set treatment $A$ to $a=2$ within the index population, $\E[ Y^{a=2} | S = 1]$, is identified by $\lambda \equiv \gamma_{0,2}+\gamma_{1,0}-\gamma_{0,0}$.
\end{lemma}

Under conditions for Lemmas 1 and 3, the average treatment effect comparing intervention to set treatment $A$ to $a=1$ vs. $a=2$ within the index population is immediate.

\begin{theorem}
If the conditions required for Lemmas 1 and 3 hold, the average treatment effect comparing intervention to set treatment $A$ to $a=1$ versus $a=2$ within the index population, $\E[ Y^{a=1} - Y^{a=2} | S = 1]$, is identified by $\phi\equiv\gamma_{1,1}-\lambda = \left(\gamma_{1,1}-\gamma_{1,0}\right)-\left(\gamma_{0,2}-\gamma_{0,0}\right).$
\end{theorem}

The statistical estimand $\phi$ represents the difference of two average treatment effects comparing the index trial with the external data, marginalized to the covariate distribution in the index trial population. This identifying functional parallels those that are of interest when pursuing ``difference-in-differences'' approaches that are popular in econometrics and the social sciences \cite{athey2006identification, goodman2021difference}. Furthermore, under this identification strategy there is no testable restriction on the law of the observed data with respect to the conditional outcome mean under intervention to set treatment $A$ to $a=0$. That is, one would not necessarily need to test in their data that for $f(x|S=s)>0$, $\E[Y|X,S=1,A=0]=\E[Y|X,S=0,A=0]$.

\section{ESTIMATION}
\label{s:estimation}

In this section, we propose estimators for $\psi$ and $\phi$, the two observed data functionals obtained by pursuing identification under transportability in mean and in effect measure, respectively. In the main text, we focus on the \textit{augmented weighting} estimators $\widehat{\psi}_{\text{AW1}}$ and $\widehat{\phi}_{\text{AW1}}$ implied by their unique influence functions under the nonparametric model, which places no restrictions on the family of probability distributions generating the observed data \cite{hampel1974influence, bickel1993efficient, robins1994estimation, rotnitzky1998semiparametric, bang2005, tsiatis2007}. Combining information from diverse sources typically involves high-dimensional settings that require components of the joint data distribution to be modeled parametrically. Here, we provide theoretical derivations to show that the augmented weighting estimators are model multiply robust in that they will be consistent for their corresponding target parameters $\psi$ and $\phi$ even if one or more of their nuisance models -- that is, for the probability of index trial participation, probability of treatment, and conditional outcome means -- are misspecified.

For completeness, we use the unique influence functions for $\psi$ and $\phi$ to define other regular and asymptotically linear estimators, provided in Sections 5 and 6 of the Supplement. These include the ``g-formula'' \cite{robins1986} estimator that uses outcome modeling and standardization (OM); weighting by the odds of index trial participation and inverse probability of treatment, using non-normalized (W1) and normalized (W2) weights; as well as two other variants of augmented weighting, specifically with normalized weights (AW2) and weighted outcome regression (AW3). We defer all proofs and most technical arguments to the Supplement.

\subsection{First order influence function for $\gamma_{s,a}$}

Recall that the statistical estimands $\psi$ and $\phi$ are linear combinations of the functionals $$\gamma_{s,a}\equiv\E\big[\E[Y|X,S=s,A=a]\big|S=1\big]$$ for all $s,a$. Therefore, our approach to estimating $\psi$ and $\phi$ will be to first derive the unique influence function for $\gamma_{s,a}$ under the nonparametric model $\mathcal{M}_{np}$ \cite{bickel1993efficient, tsiatis2007}. For notational convenience, we define the following ``true'' nuisance functions: the marginal probability of participation in the index trial, $\zeta=\Pr[S=1]$; the conditional probability of participation in a given data source, $p_s(X)=\Pr[S=s|X]$, such that for $S=1$, $p_1(X)\equiv p(X)=\Pr[S=1|X]$, and for $S=0$, $p_0(X)=1-p_1(X)=1-p(X)$; the conditional probability of treatment, $e_{s,a}(X)=\Pr[A=a|X, S=s]$, for all $s,a$; and the conditional outcome mean $g_{s,a}(X)=\E[Y|X, S=s, A=a]$ for all $s,a$. We represent these nuisance functions in the vector $\boldsymbol{\eta}_{s,a}=\big\{\zeta, p_s(X), e_{s,a}(X), g_{s,a}(X)\big\}$ for all $s,a$.

The influence function for $\gamma_{s,a}$ can be obtained by pathwise differentiability under the nonparametric model $\mathcal{M}_{np}$ \cite{bickel1993efficient, tsiatis2007}. Briefly, the first partial derivative of $\gamma_{s,a}$ is taken with respect to $t$ under the parametric submodel $\big\{\mathcal{P}_{t}:t\in[0,1)\big\}$, and evaluated at the true data law at $t=0$. The influence function for $\gamma_{s,a}$ is the unique mean-zero, finite variance random variable that satisfies the identity $\dfrac{\partial \gamma_{\mathcal{P}_t}}{\partial t}\biggr|_{t=0}=\E_{\mathcal{P}_{t=0}}\left[\Gamma(O;\boldsymbol{\eta}_{s,a})u(O)\right]$, where $u(O)$ is the full data score evaluated at $t=0$. Here, the unique influence function for $\gamma_{s,a}$ is expressed as 
$$    \Gamma_{s, a}(O; \boldsymbol{\eta}_{s,a}) =\dfrac{1}{\zeta}\left\{I(S=1)\{g_{s, a}(X)-\gamma_{s, a}\}+\dfrac{I(S=s,A=a)p_1(X)}{e_{s, a}(X)p_s(X)}\{Y-g_{s, a}(X)\}\right\},$$
for all $s,a$, with the functions $\gamma_{s,a}$, $p_s(X)$, $e_{s,a}(X)$, and $g_{s,a}(X)$ defined as above. 

\subsection{Estimation under transportability in mean} 

\subsubsection{First order influence function for $\psi$} The influence function for $\psi$, $\Psi(O;\boldsymbol{\beta})$, is the unique mean-zero, finite variance random variable that satisfies the identity $\dfrac{\partial \psi_{\mathcal{P}_t}}{\partial t}\biggr|_{t=0}=\E_{\mathcal{P}_{t=0}}[\Psi(O;\boldsymbol{\beta})u(O)].$
Using the previous results for $ \Gamma_{s, a}(O; \boldsymbol{\eta}_{s,a})$, it is straightforward to see that the unique influence function for $\psi$ under the nonparametric model $\mathcal{M}_{np}$ can be expressed as 
\begin{align*}\Psi(O;\boldsymbol{\beta})=& \dfrac{1}{\zeta}\Biggr\{I(S=1)\big\{g_{1,1}(X)-\gamma_{1,1}\big\}+\dfrac{I(S=1,A=1)}{e_{1,1}(X)} \big \{Y-g_{1,1}(X)\big\}\Biggr\}\\
    &\quad \quad \quad \quad - \dfrac{1}{\zeta}\Biggr\{I(S=1)\big\{g_{0,2}(X)-\gamma_{0,2}\big\}+\dfrac{
I(S=0,A=2)p(X)}{e_{0,2}(X)\{1-p(X)\}}\big \{Y-g_{0,2}(X)\big\}\Biggr\},
\end{align*} 

\medskip where probabilities are taken with respect to the true underlying distribution $\mathcal{P}_0$, and the nuisance functions in $\boldsymbol{\beta}=\big\{\zeta, p(X), e_{1, 1}(X), e_{0, 2}(X), g_{1, 1}(X), g_{0, 2}(X)\big\}$ defined as above.   

In Section 4 of the Supplement, we show that the unique influence function for $\psi$ under $\mathcal{M}_{np}$ is the \textit{efficient} influence function under the semiparametric model $\mathcal{M}_{sp}$ in which the probabilities of treatment $e_{s,a}(X)$ are known. This result suggests that the augmented weighting estimator constructed using $\Psi(O;\mathbf{\beta})$, provided in the next section, will have the same properties under $\mathcal{M}_{np}$ and $\mathcal{M}_{sp}$ when the propensity scores in the index and external data are known, for example when data are drawn from marginally or conditionally randomized trials.

\subsubsection{Proposed augmented weighting estimator for $\psi$} The unique influence function of $\psi$ suggests the estimator $\widehat{\psi}_{\text{AW1}}\equiv\widehat{\gamma}_{1,1}^\text{AW1}-\widehat{\gamma}_{0,2}^\text{AW1}$, where
\begin{align*}
\widehat \gamma_{s,a} ^{\text{AW1}}=\Biggr\{\sum_{i=1}^n I(S_i=1)\Biggr\}^{-1}\sum_{i=1}^{n}{\Biggr\{}I(S_i=1){\widehat{g}}_{s,a}\left(X_i\right)+\widehat{w}_{s,a}(X_i,S_i,A_i)\left\{Y_i-{\widehat{g}}_{s,a}(X_i)\right\}\Biggr\} 
\end{align*} 

\noindent for $s,a \in \{(1,1),(0,2)\}$. Here, $\widehat{g}_{s,a}(X)$ is an estimator for $g_{s,a}(X)=\E[Y|X,S=s,A=a]$; and the weight $\widehat{w}_{s,a}(X_i,S_i,A_i)$ for the $i^{th}$ observation is defined as 
$$\widehat{w}_{s,a}(X_i,S_i,A_i) = \dfrac{I(S_i=1,A_i=a)}{\widehat{e}_{1,a}(X_i)}+\dfrac{I(S_i=0,A_i=a)\widehat{p}(X_i)}{\widehat{e}_{0,a}(X_i)\left\{1-\widehat{p}(X_i)\right\}},$$   
where ${\widehat{p}}(X)$ is an estimator for $p(X)=\Pr[S=1|X]$ and ${\widehat{e}}_{s,a}(X)$ an estimator for $e_{s,a}(X)=\Pr[A=a|X,S=s]$.

Informally, $\widehat \psi_{\text{AW1}}$ corrects outcome predictions from $\widehat g_{1,1}$ and $\widehat g_{0,2}$ by weighting the residuals $Y_i-\widehat{g}_{s,a}$ by their respective weights $\widehat{w}_{1,1}$ and $\widehat{w}_{0,2}$. Thus, the construction of $\widehat \psi_{\text{AW1}}$ based on its efficient influence function is an attempt to de-bias the plugin g-formula estimator through which one obtains predictions by modeling conditional outcome means \cite{kennedy2024semiparametric}. Furthermore, under our setup, the probabilities for treatment in the index and external trials are known by design and can be estimated nonparametrically. However, modeling the probability of treatment may yield efficiency gains by adjusting for random covariate imbalances between treatment groups, separately within the index and external populations \cite{kennedy2017semiparametric}.

\subsubsection{Asymptotic properties of the augmented weighting estimator $\widehat \psi_{\text{AW1}}$} 

We now examine the asymptotic properties of the augmented weighting estimator $\widehat \psi_{\text{AW1}}$. To do so, we assume that the estimators $\widehat p(X)$, $\widehat e_{s,a}(X)$, and $\widehat g_{s,a}(X)$ have asymptotic limits $p^*(X)$, $e_{s,a}^*(X)$, and $g_{s,a}^*(X)$, respectively. We say that the estimators for nuisance functions are consistent if their asymptotic limits coincide with their true nuisance functions $p(X)$, $e_{s,a}(X)$, and $g_{s,a}(X)$, as would occur if their models used in estimation were correctly specified.

\begin{theorem} Under conditions listed under Section 5.3.2 of the Supplement, $\widehat \psi_{\text{AW1}}$ has the following asymptotic properties.\\
\textbf{(a) Consistency.} $\widehat \psi_{\text{AW1}}$ is model multiply robust in that if at least one element in each of the following sets of true nuisance functions, $\big\{e_{1, 1}(X), g_{1, 1}(X)\big\} \text{ and } \big\{\{p(X), e_{0,2}(X)\},g_{0,2}(X)\big\}$ is correctly specified, $\widehat \psi_{\text{AW1}}$ will be consistent for $\psi$. Furthermore, under rate and smoothness conditions, $\widehat \psi_{\text{AW1}}$ will be rate robust in that it will be consistent at parametric, or $\sqrt{n}$, rate. \\
\textbf{(b) Asymptotic normality.} The augmented weighting estimator $\widehat \psi_{\text{AW1}}$ centered at $\psi$ and scaled by $\sqrt{n}$ converges to a normal distribution with mean 0 and asymptotic variance equal to the variance of the influence function $\Psi\left(O; \boldsymbol{\beta}\right)$. That is, $$\sqrt{n}(\widehat \psi_{\text{AW1}}-\psi) 
\xrightarrow{D} \mathcal{N}\left(0,\E\left[\big(\Psi(O; \boldsymbol{\beta})\big)^2\right]\right).$$
\end{theorem}
Since nonparametric estimates of the propensity scores $e_{1,1}(X)$ and $e_{0,2}(X)$ are known to investigators, for $\widehat \psi_{\text{AW1}}$ to be consistent it is sufficient for the model for the conditional probability of trial participation, $\widehat p(X)$, or the model for either the conditional outcome mean in the external data, $\widehat g_{0,2}(X)$, to be correctly specified. Furthermore, the rate robustness properties of $\widehat \psi_{\text{AW1}}$ allow for slower convergence rates for some nuisance models (e.g., using data-adaptive methods), provided the other function(s) in their same sets converge at rates that can compensate in a manner that allows $\widehat \psi_{\text{AW1}}$ to still achieve $\sqrt{n}$-consistency \cite{chernozhukov2017double, chernozhukov2018double}.

\subsection{Estimation under transportability in effect measure}

\subsubsection{First order influence function for $\phi$} The influence function for $\phi$, $\Phi(O;\boldsymbol{\tau})$, is the unique mean-zero, finite variance random variable that satisfies the identity $\dfrac{\partial \phi_{\mathcal{P}_t}}{\partial t}\biggr|_{t=0}=\E_{\mathcal{P}_{t=0}}[\Phi(O;\boldsymbol{\tau})u(O)].$ Using earlier results for $\Gamma_{s,a}$, the unique influence function for $\phi$ under the nonparametric model can be expressed as $$\Phi(O;\boldsymbol{\tau})=\Gamma_{1, 1}(O; \boldsymbol{\eta}_{1,1})-\Gamma_{1, 0}(O; \boldsymbol{\eta}_{1,0})-\Gamma_{0, 2}(O; \boldsymbol{\eta}_{0,2})+\Gamma_{0, 0}(O; \boldsymbol{\eta}_{0,0}),$$
where $\boldsymbol{\eta}_{s,a}$ is defined as above; $\boldsymbol{\tau}=\big\{\zeta, p(X), \textbf{e}_{s, a}(X), \textbf{g}_{s, a}(X)\big\}$ includes the family of functions $\textbf{e}_{s, a}(X)$ and $\textbf{g}_{s, a}(X)$ for all $s,a$; and all probabilities are taken with respect to the true underlying data-generating distribution $\mathcal{P}_0$. More explicitly, $\Phi(O;\boldsymbol{\tau})$ can be written using
\noindent\resizebox{\columnwidth}{!}{%
\begin{minipage}{\columnwidth}
\begin{align*}
\Gamma_{1, a}(O; \boldsymbol{\eta}_{1,a}) = \dfrac{1}{\zeta}\left\{I(S=1)\{g_{1, a}(X)-\gamma_{1, a}\}+\dfrac{I(S=1,A=a)}{e_{1, a}(X)}\{Y-g_{1, a}(X)\}\right\} \text{for\;} a=0,1 \text{\;and;}
\end{align*}
\end{minipage}} 
\noindent\resizebox{\columnwidth}{!}{%
\begin{minipage}{\columnwidth}
\begin{align*}
\Gamma_{0, a}(O; \boldsymbol{\eta}_{0,a}) = \dfrac{1}{\zeta}\left\{I(S=1)\{g_{0, a}(X)-\gamma_{0, a}\}+\dfrac{I(S=0,A=a)p(X)}{e_{0, a}(X)\{1-p(X)\}}\{Y-g_{0, a}(X)\}\right\} \text{for\;} a=0,2. 
\end{align*}
\end{minipage}} \newline

In Section 6 of the Supplement, we also show that $\Phi(O; \boldsymbol{\tau})$ is the \textit{efficient} influence function under the semiparametric model $\mathcal{M}_{sp}$ where the probabilities of treatment $e_{s,a}(X)$ are known for all $s,a$. This suggests that the augmented weighting estimator implied by $\Phi(O;\boldsymbol{\tau})$, provided in the next section, will have the same properties under $\mathcal{M}_{np}$ and $\mathcal{M}_{sp}$ if the propensity scores in the index and external populations are known.

\subsubsection{Proposed augmented weighting estimator for $\phi$} The influence function for $\phi$ suggests the estimator
$${\widehat{\phi}}_{\text{AW1}}= \widehat \gamma_{1,1}^{\text{AW1}} - \widehat \gamma_{1,0}^{\text{AW1}} - \left(\widehat \gamma_{0,2}^{\text{AW1}} - \widehat \gamma_{0,0}^{\text{AW1}}\right),$$ using the same definition of $\widehat \gamma_{s,a} ^{\text{AW1}}$ provided in the previous section.  

\subsubsection{Asymptotic properties of the augmented weighting estimator $\widehat \phi_{\text{AW1}}$}

In this section, we examine the asymptotic properties of $\widehat \phi_{\text{AW1}}$ based on its asymptotic representation, which we provide in Section 6 of the Supplement alongside needed technical arguments.

\begin{theorem} Under technical conditions listed in Section 6.3.2 of the Supplement,  $\widehat \phi_{\text{AW1}}$ has the following asymptotic properties.\mbox{}\\
\textbf{(a) Consistency.} $\widehat \phi_{\text{AW1}}$ is model multiply robust in the sense that if at least one element in each of the following sets of nuisance functions, $\{e_{1, 1}(X),g_{1, 1}(X)\}$, $\{e_{1, 0}(X),g_{1, 0}(X)\}$, \linebreak $\big\{\left\{p(X), e_{0,2} (X)\right\},g_{0,2}(X)\big\}$, \text{and} $\big\{\{p(X), e_{0,0} (X)\},g_{0, 0}(X)\big\}$ is correctly specified, $\widehat \phi_{\text{AW1}}$ will be consistent for $\phi$. Furthermore, under rate and smoothness conditions, $\widehat \phi_{\text{AW1}}$ will be rate robust in that it will be consistent for $\phi$ at parametric, or $\sqrt{n}$, rate.\\
\noindent \textbf{(b) Asymptotic normality.} The augmented weighting estimator $\widehat \phi_{\text{AW1}}$ centered at $\phi$ and scaled by $\sqrt{n}$ converges to a normal distribution with mean 0 and asymptotic variance equal to the variance of the influence function $\Phi(O;\boldsymbol{\tau})$. That is, $$\sqrt{n}\left(\widehat{\phi}_{\text{AW1}}-\phi\right) \xrightarrow{D} \mathcal{N}\left(0,\E\left[\big(\Phi(O;\boldsymbol{\tau})\big)^2\right]\right).$$
\end{theorem}

\subsection{Comparing the efficiency of $\widehat \psi_{\text{AW1}}$ and $\widehat \phi_{\text{AW1}}$} In this section, we compare the asymptotic efficiencies for $\widehat \psi_{\text{AW1}}$ and $\widehat \phi_{\text{AW1}}$, assuming that the conditions under which they are consistent and asymptotically normal hold. This comparison may be of interest because $\E[Y^{a=1}-Y^{a=2}|S=1]$ can be identified with $\psi$ \textit{and} $\phi$ under conditional mean exchangeability over trial participation $S$ (A4), with respect to intervention to set treatment $A$ to $a=0,2$. More specifically, if A4 holds for $a=0$, a condition that is not strictly required for identification under transportability in mean (see Lemma 2 and Theorem 1), then the observed data functional $\psi$ can be viewed as a special case of $\phi$. To see this, note that condition A4 for $a=0$ requires $\E[Y^{a=0}|X,S=1]=\E[Y^{a=0}|X,S=0]$. This condition, together with exchangeability over treatment (A2 and A6) and consistency (A1), imposes the testable restriction $\E[Y|X,S=1,A=0]=\E[Y|X,S=0,A=0]$ on the observed data.

To compare the efficiency of $\widehat \psi_{\text{AW1}}$ and $\widehat \phi_{\text{AW1}}$ under this scenario, we describe two intermediate results. Full proofs can be found in Section 7 of the Supplement. First, we introduce the function 
$\delta \equiv \phi - \psi$. Using the result for $\Gamma_{s,a}$, the influence function for $\delta$ under the nonparametric model is given by $\Delta(O;\boldsymbol{\upsilon})=\Gamma_{1,0}(O;\boldsymbol{\eta}_{1,0})-\Gamma_{0,0}(O;\boldsymbol{\eta}_{0,0})$, where $\boldsymbol{\upsilon}=\{\zeta, p(X), e_{1,0}(X),e_{0,0}(X),g_{1,0}(X),g_{0,0}(X)\}$. The augmented weighting estimator for $\delta$ can be defined as $\widehat \delta_{\text{AW1}}=\widehat \gamma_{1,0}^{\text{AW1}} - \widehat\gamma_{0,0}^{\text{AW1}}$, and its asymptotic distribution is immediate using arguments symmetrical to those presented in Theorems 3 and 4. 
\begin{lemma} The asymptotic distribution of the augmented weighting estimator $\widehat \delta_{\text{AW1}}$ is
 $$\sqrt{n}(\widehat \delta_{\text{AW1}}-\delta) \xrightarrow{D} \mathcal{N}\left(0,\E\left[\big(\Delta(O; \boldsymbol{\upsilon})\big)^2\right]\right),$$
where the mean and asymptotic variance equals the mean and variance of the influence function $\Delta(O; \boldsymbol{\upsilon})$.
\end{lemma}

Secondly, we examine the asymptotic covariance of the influence functions for $\psi$ and $\delta$.

\begin{lemma} Taken with respect to the combined data distribution $O=(X,S,A,Y)$, the expectation of the product of the influence functions $\Psi(O;\boldsymbol{\beta})$ and $\Delta(O;\boldsymbol{\upsilon})$ for $\psi$ and $\delta$, respectively, is 0. That is, $\E\big[\Delta(O;\boldsymbol{\upsilon})\Psi(O;\boldsymbol{\beta})\big]=0$. This implies that the asymptotic covariance of the augmented weighting estimators $\widehat \psi_{\text{AW1}}$ and $\widehat \delta_{\text{AW1}}$ is also 0.
\end{lemma}

This result only holds if there is transportability in mean (A4) for intervention to set treatment $A$ to $a=0$. Using the results of Lemmas 4 and 5, we can now express the efficiency bound of $\widehat \phi_{\text{AW1}}$ by summing the asymptotic variances of $\widehat \psi_{\text{AW1}}$ and $\widehat \delta_{\text{AW1}}$.

\begin{theorem} Under technical conditions listed in Sections 5.3.2 and 6.3.2 of the Supplement, and using the results of Lemmas 4 and 5, the asymptotic efficiency gain of using the augmented weighting estimator $\widehat \psi_{\text{AW1}}$ instead of $\widehat \phi_{\text{AW1}}$ to estimate $\E[Y^{a=1}-Y^{a=2}|S=1]$ is given by $$\E\left[\big(\Phi(O;\boldsymbol{\tau}\big)^2\right] - \E\left[\big(\Psi(O;\boldsymbol{\beta})\big)^2\right]= \E\left[\big(\Delta(O;\boldsymbol{\upsilon})\big)^2\right]>0,$$ 
where
$\E\left[\big(\Phi(O;\boldsymbol{\tau})\big)^2\right]$, $\E\left[\big(\Psi(O;\boldsymbol{\beta})\big)^2\right]$, and $\E\left[\big(\Delta(O;\boldsymbol{\upsilon})\big)^2\right]$ are the asymptotic variances of the augmented weighting estimators $\widehat \phi_{\text{AW1}}$, $\widehat \psi_{\text{AW1}}$, and $\widehat \delta_{\text{AW1}}$, respectively. 
\end{theorem}

These results suggest that when transportability in mean (A4) holds for intervention to set treatment $A$ to $a=2$ \textit{and} $a=0$, both $\widehat \psi_{\text{AW1}}$ and $\widehat \phi_{\text{AW1}}$ will provide consistent estimates for the same causal parameter, but $\widehat \phi_{\text{AW1}}$ will be less efficient because of the need to fit additional nuisance models $\widehat g_{1,0}$ and $\widehat g_{0,0}$. Finally, we note that these results will also pertain to scenarios where there is conditional mean exchangeability over $S$ for $a=2$, and conditional exchangeability (transportability) over $S$ for mean differences for $a=2,0$; these two conditions together imply that there is conditional mean exchangeability over $S$ for $a=0$ (see Section 7 of the Supplement for further details). In practice, the decision to choose between $\widehat \psi_{\text{AW1}}$ and $\widehat \phi_{\text{AW1}}$ may be informed by examining the testable restriction $\E[Y|X,S=1,A=0]=\E[Y|X,S=0,A=0]$, bearing in mind well-known limitations of such falsification tests \cite{bancroft1944biases, giles1993pre, dahabreh2023using}.  

\subsection{Statistical inference} Consistent estimation of sampling variances for the estimators presented in this report may be obtained with robust standard errors as in M-estimation \cite{stefanski2002} or nonparametric bootstrapping \cite{efron1994introduction}. For the proposed augmented weighting estimators, both methods will yield consistent variance estimates as long as a sufficient set of nuisance models needed for consistent point estimation is correctly specified \cite{shook2025double, wu2025variance}.

\section{SIMULATION STUDY}
\label{s:simulation}

We conducted a simulation study, modeled closely after \cite{dahabreh2023efficient}, to evaluate the finite sample performance of augmented weighting estimators (AW1) for $\psi$ and $\phi$. For completeness, we also included OM, W1, W2, AW2, and AW3 estimators. Full methods and results are available in Section 8 of the Supplement.

\subsection{Data generation} We combined index and external data drawn from marginally randomized trials: an index trial examining $A=1$ vs. $A=0$, and an external trial examining $A=2$ vs. $A=0$. Briefly, the data generation process involved the following steps:\\

\noindent \textbf{(1)} \textit{Underlying superpopulation}. A population of size $n=10^6$ was generated, with independent and identically distributed observations $(X_{1i},X_{2i},X_{3i})$ following a multivariate normal distribution with marginal means 0, marginal variances 1, and pairwise covariances 0.5.\\

\noindent \textbf{(2)} \textit{Selection for participation in a trial.} Individuals from the underlying superpopulation were randomly chosen to participate in either of the index or external trials, under different total sample size scenarios where $n=1,000$, $n=2,000$, and $n=10,000$.\\

\noindent \textbf{(3)} \textit{Allocation of individuals into the index and external trials.} To allocate individuals into either the index or external trials, we fit a logistic regression for the probability of index trial participation, $\Pr[S=1|\boldsymbol{X}]=\dfrac{\text{exp}{(\boldsymbol{\alpha}^\top\boldsymbol{X}})}{1+\text{exp}{(\boldsymbol{\alpha}^\top\boldsymbol{X}})}$, where $\boldsymbol{X}=\{1,X_1,X_2,X_3\}$ and $\boldsymbol{\alpha}=\{\alpha_0,\ln 2,\ln 2,\ln 2\}$. Numerical methods were used to set the intercept $\alpha_0$ to values that would result in the desired sample sizes in the index and external trials \cite{robertson2022using}.\\

\noindent \textbf{(4)} \textit{Random treatment assignment.} Independent Bernoulli random variables $A_i$ with parameters $\mathrm{Pr}[A=a|S=s]=0.5$ were generated for the index and external trials.\\

\noindent \textbf{(5)} \textit{Potential and observed outcomes.} For each individual, we generated potential outcomes using the model $Y_i^a=(\boldsymbol{\upsilon}^a)^\top\boldsymbol{X}+\epsilon$, where $\boldsymbol\upsilon^{a=1}=\boldsymbol\upsilon^{a=2}=(\upsilon_0, 1,1,1)$, $\boldsymbol\upsilon^{a=0}=(\upsilon_0,-1,-1,-1)$, and $\epsilon \sim \mathcal{N}(0,1)$. The coefficients $\boldsymbol\upsilon^a$ were chosen such that $\E[Y^{a=1}-Y^{a=2}|S=1]=0$, and under conditional mean exchangeability over $S$, that is $\E[Y^a|X,S=1]=\E[Y^a|X,S=0]$ for $a=0,1,2$. Observed outcomes were generated under consistency, where $Y_i^{a}=Y_i$ for $A_i=a$.

\subsection{Scenarios and performance metrics} We conducted simulations under the following sample size scenarios: (1) $n_{S=1}=500, n_{S=0}=500$; (2) $n_{S=1}=800, n_{S=0}=200$; (3) $n_{S=1}=200, n_{S=0}=800$; (4) $n_{S=1}=1,000, n_{S=0}=1,000$, and (5) $n_{S=1}=5,000, n_{S=0}=5,000$. Simulations were conducted under correct model specification for nuisance functions, as well as three scenarios under which these models were misspecified: (1) misspecification of models for $p(X)$ and $e_{s,a}(X)$, corresponding to the probability of index trial participation and of treatment within each trial, respectively; (2) misspecification of models for $g_{s,a}(X)$, corresponding to conditional outcome means; (3) misspecification of all models. Misspecified models included only an $\epsilon \sim \mathcal{N}(0,1)$ error term and an intercept. Each scenario was run for 10,000 iterations. For each estimator, we reported the mean bias, empirical standard errors, and the mean squared error, defined as the sum of the squared mean bias and variance of each estimator obtained over all iterations. Detailed methods for these performance metrics can be found in Section 8 of the Supplement.

\subsection{Results} Tables \ref{tab:simulation_correct} and \ref{tab:simulation_misspecified} provide simulation results under different sample size and model specification scenarios. When all models were specified correctly, $\widehat{\psi}_{\text{AW1}}$ and $\widehat{\phi}_{\text{AW1}}$ had negligible bias across all sample size scenarios. The empirical standard errors for all estimators of $\psi$ were uniformly lower than the corresponding estimators for $\phi$. Furthermore, as anticipated by our theoretical derivations, the augmented weighting estimators remained consistent for $\psi$ and $\phi$ even under misspecification of the models for the probability of index trial participation $p(X)$ and for treatment $e_{s,a}(X)$; and, separately, for the conditional outcome means, $g_{s,a}(X)$.

\section{CLINICAL APPLICATION}
\label{s:application}

\subsection{Data description} ACCEPT $(2007 - 2009) $\cite{griffiths2010comparison} and PHOENIX  $(2005 - 2011) $\cite{leonardi2008efficacy} were double-blinded randomized trials that evaluated biologic agents to treat moderate-to-severe plaque psoriasis, a common systemic inflammatory disorder characterized by scaly erythematous lesions of the skin, notable extracutaneous sequelae (e.g., cardiovascular, renal, and joint involvement), effects on mental health, and diminished quality of life \cite{gelfand2006risk, griffiths2021psoriasis}. Both trials enrolled adults aged $\geq 18$ with a baseline psoriasis activity and severity index (PASI) score of $\geq 12$ and involvement of $\geq 10\%$ body surface area. The ACCEPT trial randomized $903$ participants to the tumor necrosis factor inhibitor etanercept 50mg or the interleukin-23 receptor inhibitor ustekinumab in 45mg or 90mg formulations, with each ustekinumab dosage considered its own treatment group. The PHOENIX trial randomized $766$ participants to ustekinumab in 45mg or 90mg formulations, or placebo. The primary end point in both trials was the proportion of participants who achieved a $\geq75\%$ reduction in their PASI (PASI 75) comparing baseline and week 12 post-randomization.

The data structure for our analyses is shown in Table \ref{tab:accept_phoenix_datastructure}, and baseline covariates of participants shown in Table \ref{tab:accept_phoenix_tab1}. We ignored the small missingness in baseline covariates or the outcome that was seen in $2.7\%$ and $1.0\%$ of participants in the ACCEPT and PHOENIX trials, respectively. Data were available through the Yale University Open Data Access Project \cite{krumholz2016yale} and our analysis approved by our local Institutional Review Board. A formalization of study design using the target trial framework \cite{hernan2016using, hernan2025target} is offered in Section 9 of the Supplement. 

\subsection{Causal estimands} We wish to identify the probability of achieving PASI 75 at 12 weeks, comparing randomization to etanercept versus placebo, within the population underlying the ACCEPT trial. 

\subsection{Identification} We pursue identification under three main assumptions. First, we assume that trial engagement effects were negligible, which may be a reasonable assumption because the ACCEPT and PHOENIX trials were conducted with near-identical protocols and follow-up procedures. Second, we assume conditional exchangeability (transportability) between the ACCEPT and PHOENIX trial participants, within levels of baseline covariates. These covariates included age; sex; percentage body surface area affected; PASI; psoriatic arthritis; ever-smoking; diabetes; and previous use of phototherapy, systemic agents, and biologic agents. This transportability condition is assumed to hold for the mean (for $\psi$), and in difference effect measure, using ustekinumab 45mg ($\phi_1$) or 90mg ($\phi_2$) as the common comparators between both trials. Third, we assume positivity in that the covariate patterns important for the transportability assumptions, either in mean or in mean difference, have common support between the populations ACCEPT and PHOENIX.

\subsection{Estimation} We implemented the AW1, AW2, AW3, OM, W1, and W2 estimators for $\psi$, $\phi_1$, and $\phi_2$. The conditional probability of trial participation was modeled with logistic regression fit within the entire composite dataset, and the conditional probability of treatment was modeled with multinomial regression separately within each trial. Conditional outcome means were estimated using logistic regression fit within each trial and treatment strata to account for all treatment by covariate product terms. All models used the following baseline covariates in their stated functional forms: age in continuous years (linear), sex (binary), percentage body surface area affected (linear), PASI (linear), psoriatic arthritis (binary), ever-smoking (binary), diabetes mellitus (binary), and previous use of phototherapy, systemic agents, and biologic agents (all binary). Nonparametric bootstrapping with 10,000 resamples was implemented to obtain percentiles-based $95\%$ confidence intervals. Analyses were performed in R version 4.4.1.

\subsection{Results} Selected results are presented in Table \ref{tab:accept_phoenix_causalcontrast}. All estimators for $\psi$, $\phi_1$ and $\phi_2$ yielded around a $50$ percentage point difference in the probability of achieving PASI 75 at 12 weeks comparing etanercept to placebo, within the population underlying the ACCEPT trial. For each estimator for $\psi$, $\phi_1$ and $\phi_2$, standard errors for $\widehat \psi$ were uniformly lower in magnitude than for $\widehat \phi_1$ and $\widehat \phi_2$. Full results of the clinical application are found in Section 9 of the Supplement.

\section{DISCUSSION}
\label{s:discussion}

We propose causal and statistical methods for combining information to construct external comparators in a time-fixed setting with point treatments and non-failure time outcomes. In practice, these identification strategies require iterative appraisals a reasonable sampling model, causal model, causal estimands, and identifiability conditions that justify the use of external data to learn about target populations and their subsets \cite{ung2025keep}. A natural organization of identification strategies is by whether transportability in mean or effect measure, or both, are believed to hold in view of domain expertise of the causal problem. 

For both identification strategies, we derived a comprehensive suite of estimators that use well-known principles of semiparametric efficiency theory \cite{robins1994estimation, robins1997marginal, rotnitzky1998semiparametric, bang2005}. We highlighted augmented weighting estimators that are semiparametric efficient when the probability of treatment in the index and external data are known; have good properties asymptotically and in finite samples; and which are particularly appealing in high-dimensional data-adaptive settings, given their model and rate robustness properties \cite{chernozhukov2017double, chernozhukov2018double}. However, the other estimators provided in Sections 5 and 6 Supplement, including those based on outcome modeling and weighting alone, should provide researchers with diverse and reasonably familiar estimation procedures that are easily implemented using standard statistical software. We applied these estimation procedures to a clinical study combining the ACCEPT and PHOENIX trials under ideal conditions -- including aligned eligibility criteria, treatments, and outcomes; near complete follow-up, and nominal missingness. The magnitude of causal effects in our study were similar across all estimation procedures, under different transportability assumptions, and corroborated by 
and a prior landmark trial that demonstrated a striking therapeutic benefit of etanercept over placebo \cite{leonardi2003etanercept}. 

Future directions for this research include extensions to more complex data structures, for example when considering failure-time outcomes and censoring. Furthermore, in applied settings there will be a need to more deeply examine study design challenges that can arise with imperfect data alignment. Such difficulties may relate to the definition of different variables, for example when combining trial-based notions assignment with observational analogs sourced externally \cite{dahabreh2025assignment}; or when data ascertainment procedures differ between data sources. Nevertheless, it is hoped that formalizing causal and statistical approaches for external comparator analyses may expand opportunities to use, in a principled manner, ever-increasing volumes of data generated in the health sciences to answer causal questions.




\newpage
\section{TABLES}
\begin{table}[H]
\centering 
\renewcommand{\arraystretch}{1}
\caption{Schematic representation of the data structure under consideration in this report, where $X$ is a vector of baseline covariates; $S$ is a binary indicator for participation within the index trial ($S=1$) or in the external data ($S=0$); $A$ is an indicator for the treatment assignment; and $Y$ is a non-failure time outcome measured at the end of follow-up.}
\renewcommand{\arraystretch}{1.3}
\label{tab:datastructure}
\resizebox{1\columnwidth}{!}{%
\begin{tabular}{|c|c|c|c|}
\hline
\textbf{Baseline covariates $(X)$} &
\textbf{Trial participation $(S)$} &
\textbf{Treatment $(A)$} &
\textbf{Outcomes $(Y)$} \\ \hline

\multirow{4}{*}{$X$} 
  & \multirow{2}{*}{$S = 1$} & $A = 1$ & \multirow{4}{*}{$Y$} \\ \cline{3-3}
  &                           & $A = 0$ &                      \\ \cline{2-3}
  & \multirow{2}{*}{$S = 0$} & $A = 0$ &                      \\ \cline{3-3}
  &                           & $A = 2$ &                      \\ \hline

\end{tabular}%
}
\end{table}


\begin{landscape}
\begin{table}[]
\centering 
\caption{Results of simulation experiment to estimate $\psi$ and $\phi$, under correctly specified nuisance models. We report the mean bias, standard error (SE), and mean squared error (MSE) across 10,000 iterations for different sample size scenarios; if not stated otherwise, $n_{S=1} = n_{S=0}$. OM: outcome modeling followed by standardization (``g-formula'' estimator); W1: non-normalized weighting using odds of index trial participation and inverse probability of treatment weights; W2: normalized weighting using odds of index trial participation and inverse probability of treatment weights; AW1: augmented weighting with non-normalized weights; AW2: augmented weighting with normalized weights; and AW3: augmented weighting using weighted outcome regression.}
\label{tab:simulation_correct}
\renewcommand{\arraystretch}{1.6}
\resizebox{\columnwidth}{!}{%
\begin{tabular}{l|ccccccccccccccc}
\hline
\multicolumn{1}{c|}{\textbf{Sample sizes}} & \multicolumn{3}{c}{$n=1,000$} & \multicolumn{3}{c}{$n=1,000 \; (n_{S=1}=800)$} & \multicolumn{3}{c}{$n=1,000 \; (n_{S=1}=200)$} & \multicolumn{3}{c}{$n=2,000$} & \multicolumn{3}{c}{$n=10,000$} \\ \hline
\multicolumn{1}{c|}{\textbf{Estimators}} & \textbf{Bias} & \textbf{SE} & \textbf{MSE} & \textbf{Bias} & \textbf{SE} & \textbf{MSE} & \textbf{Bias} & \textbf{SE} & \textbf{MSE} & \textbf{Bias} & \textbf{SE} & \textbf{MSE} & \textbf{Bias} & \textbf{SE} & \textbf{MSE} \\ \hline
$\widehat\psi_{\text{OM}}$ & 0.0010 & 0.1270 & 0.0161 & 0.0014 & 0.1934 & 0.0374 & -0.0005 & 0.1321 & 0.0174 & -0.0001 & 0.0880 & 0.0077 & 0.0004 & 0.0397 & 0.0016 \\
$\widehat\psi_{\text{W1}}$ & 0.0054 & 0.8772 & 0.7694 & 0.0019 & 1.6134 & 2.6031 & 0.0066 & 0.7781 & 0.6055 & -0.0076 & 0.6196 & 0.3840 & 0.0013 & 0.2597 & 0.0675 \\
$\widehat\psi_{\text{W2}}$ & 0.0702 & 0.5174 & 0.2726 & 0.2041 & 0.7964 & 0.6759 & 0.0400 & 0.4265 & 0.1835 & 0.0317 & 0.4060 & 0.1659 & 0.0091 & 0.1969 & 0.0389 \\
$\widehat\psi_{\text{AW1}}$ & 0.0002 & 0.1979 & 0.0392 & 0.0012 & 0.3162 & 0.1000 & -0.0025 & 0.1728 & 0.0299 & -0.0006 & 0.1327 & 0.0176 & 0.0010 & 0.0585 & 0.0034 \\
$\widehat\psi_{\text{AW2}}$ & 0.0007 & 0.1715 & 0.0294 & 0.0020 & 0.2562 & 0.0657 & -0.0020 & 0.1626 & 0.0265 & -0.0004 & 0.1252 & 0.0157 & 0.0009 & 0.0577 & 0.0033 \\
$\widehat\psi_{\text{AW3}}$ & 0.0007 & 0.1642 & 0.0270 & 0.0020 & 0.2632 & 0.0693 & -0.0018 & 0.1584 & 0.0251 & -0.0003 & 0.1184 & 0.0140 & 0.0008 & 0.0555 & 0.0031 \\ \hline
$\widehat\phi_{\text{OM}}$ & 0.0006 & 0.1784 & 0.0318 & 0.0027 & 0.2718 & 0.0739 & -0.0018 & 0.1874 & 0.0351 & 0.0005 & 0.1253 & 0.0157 & 0.0002 & 0.0553 & 0.0031 \\
$\widehat\phi_{\text{W1}}$ & 0.0149 & 1.3293 & 1.7672 & 0.0259 & 2.1118 & 4.4602 & -0.0051 & 0.9144 & 0.8362 & 0.0154 & 0.7883 & 0.6217 & -0.0045 & 0.3561 & 0.1269 \\
$\widehat\phi_{\text{W2}}$ & 0.1459 & 0.6801 & 0.4839 & 0.4175 & 1.0649 & 1.3084 & 0.0674 & 0.5177 & 0.2726 & 0.0826 & 0.5252 & 0.2826 & 0.0123 & 0.2718 & 0.0740 \\
$\widehat\phi_{\text{AW1}}$ & 0.0025 & 0.2971 & 0.0883 & 0.0041 & 0.4388 & 0.1925 & -0.0049 & 0.2423 & 0.0587 & 0.0004 & 0.1856 & 0.0344 & 0.0006 & 0.0820 & 0.0067 \\
$\widehat\phi_{\text{AW2}}$ & 0.0013 & 0.2407 & 0.0579 & 0.0039 & 0.3629 & 0.1317 & -0.0042 & 0.2315 & 0.0536 & 0.0005 & 0.1755 & 0.0308 & 0.0006 & 0.0809 & 0.0066 \\
$\widehat\phi_{\text{AW3}}$ & 0.0002 & 0.2316 & 0.0536 & 0.0035 & 0.3718 & 0.1382 & -0.0038 & 0.2259 & 0.0511 & 0.0005 & 0.1675 & 0.0281 & 0.0004 & 0.0778 & 0.0061 \\ \hline
\end{tabular}%
}
\end{table}
\end{landscape}


\begin{landscape}
\begin{table}[]
\centering
\caption{Simulation results to evaluate estimators for $\psi$ and $\phi$ under misspecification of the models for (1) the probability of index trial participation $p(X)$ and the probability of treatment $e_{s,a}(X)$; (2)  conditional outcome means $g_{s,a}(X)$; and (3) all nuisance models. Models were misspecified by only including an $\epsilon \sim \mathcal{N}(0,1)$ error term and an intercept. $10,000$ simulations were conducted with a total sample size of $n=10,000$, with $n_{S=1} = n_{S=0} = 5,000$. SE: standard error; MSE: mean squared error.}\renewcommand{\arraystretch}{1.8}
\label{tab:simulation_misspecified}
\renewcommand{\arraystretch}{1}{%
\begin{tabular}{lccccccccc}
\hline
\multicolumn{1}{c}{\textbf{Model misspecification}} &
  \multicolumn{3}{c}{$\widehat p(X)$ and $\widehat e_{s,a}(X)$} &
  \multicolumn{3}{c}{$\widehat g_{s,a}(X)$} &
  \multicolumn{3}{c}{$\widehat p(X)$, $\widehat e_{s,a}(X)$ and $\widehat g_{s,a}(X)$} \\ \hline
\textbf{Estimator} &
  \textbf{Bias} &
  \textbf{SE} &
  \textbf{MSE} &
  \textbf{Bias} &
  \textbf{SE} &
  \textbf{MSE} &
  \textbf{Bias} &
  \textbf{SE} &
  \textbf{MSE} \\ \hline
$\widehat{\psi}_{\text{OM}}$  & 0.0004 & 0.0397 & 0.0016  & 2.7689  & 0.0571 & 7.6699  & 2.7689 & 0.0571 & 7.6699  \\
$\widehat{\psi}_{\text{W1}}$  & 2.7689 & 0.0571 & 7.6699  & 0.0013  & 0.2597 & 0.0675  & 2.7689 & 0.0571 & 7.6699  \\
$\widehat{\psi}_{\text{W2}}$  & 2.7689 & 0.0571 & 7.6699  & 0.0091  & 0.1969 & 0.0389  & 2.7689 & 0.0571 & 7.6699  \\
$\widehat{\psi}_{\text{AW1}}$ & 0.0003 & 0.0273 & 0.0007  & 0.0018  & 0.3137 & 0.0984  & 2.7689 & 0.0571 & 7.6699  \\
$\widehat{\psi}_{\text{AW2}}$ & 0.0003 & 0.0273 & 0.0007  & 0.0093  & 0.1901 & 0.0362  & 2.7689 & 0.0571 & 7.6699  \\
$\widehat{\psi}_{\text{AW3}}$ & 0.0003 & 0.0273 & 0.0007  & 0.0162  & 0.1838 & 0.0341  & 2.7689 & 0.0571 & 7.6699  \\ \hline
$\widehat{\phi}_{\text{OM}}$  & 0.0002 & 0.0553 & 0.0031  & 5.5370  & 0.0804 & 30.6645 & 5.5370 & 0.0804 & 30.6645 \\
$\widehat{\phi}_{\text{W1}}$  & 5.5370 & 0.0804 & 30.6645 & -0.0045 & 0.3561 & 0.1269  & 5.5370 & 0.0804 & 30.6645 \\
$\widehat{\phi}_{\text{W2}}$  & 5.5370 & 0.0804 & 30.6645 & 0.0123  & 0.2718 & 0.0740  & 5.5370 & 0.0804 & 30.6645 \\
$\widehat{\phi}_{\text{AW1}}$ & 0.0002 & 0.0384 & 0.0015  & -0.0030 & 0.4189 & 0.1755  & 5.5370 & 0.0804 & 30.6645 \\
$\widehat{\phi}_{\text{AW2}}$ & 0.0002 & 0.0384 & 0.0015  & 0.0140  & 0.2606 & 0.0681  & 5.5370 & 0.0804 & 30.6645 \\
$\widehat{\phi}_{\text{AW3}}$ & 0.0002 & 0.0384 & 0.0015  & 0.0279  & 0.2512 & 0.0639  & 5.5370 & 0.0804 & 30.6645 \\ \hline
\end{tabular}%
}
\end{table}
\end{landscape}

\begin{landscape}
\begin{table}[]
\centering
\caption{Data structure of the combined ACCEPT and PHOENIX trial populations. BSA: body surface area; PGA: physician global assessment; PASI: psoriasis activity severity index; PASI 75: $\geq 75\%$ reduction in PASI.}
\renewcommand{\arraystretch}{1.3}
\label{tab:accept_phoenix_datastructure}
\resizebox{\columnwidth}{!}{%
\begin{tabular}{|l|l|l|l|}
\hline
\multicolumn{1}{|c|}{\textbf{Baseline covariates ($X$)}} & \multicolumn{1}{c|}{\textbf{Study ($S$)}} & \multicolumn{1}{c|}{\textbf{Treatment ($A$)}} & \multicolumn{1}{c|}{\textbf{Outcome ($Y$)}} \\ \hline
\multirow{6}{*}{\begin{tabular}[c]{@{}l@{}}Age, sex, BSA affected, PASI, PGA, psoriatic arthritis, smoking status, diabetes \\ mellitus, and previous interventions (phototherapy, systemic agents, biologics)\end{tabular}} & \multirow{3}{*}{ACCEPT ($S=1$)} & $A=1$: Etanercept & \multirow{6}{*}{PASI 75 at 12 weeks} \\ \cline{3-3}
 &  & $A=0$: Ustekinumab 45mg &  \\ \cline{3-3}
 &  & $A=0^*$: Ustekinumab 90mg &  \\ \cline{2-3}
 & \multirow{3}{*}{PHOENIX ($S=0$)} & $A=0$: Ustekinumab 45mg &  \\ \cline{3-3}
 &  & $A=0^*$: Ustekinumab 90mg &  \\ \cline{3-3}
 &  & $A=2$: Placebo &  \\ \hline
\end{tabular}%
}
\end{table}
\end{landscape}

\begin{landscape}
\begin{table}[]
\centering
\caption{Characteristics of ACCEPT and PHOENIX trial participants, stratified by treatment assignment, and including missing data. Measures of age in the ACCEPT trial were transformed for anonymization procedures prior to data release: 60\% was generalized to 1-year; 4\% to 2 years; 4\% to 5 years, and 32\% replaced with randomly chosen value. There were $n=25$ and $n=8$ individuals from the ACCEPT and PHOENIX trials, respectively, who had at least one missing value across baseline covariates or the outcome of PASI 75 at 12 weeks following randomization. These individuals were provisionally ignored such that the complete cases numbered $n=878$ and $n=758$ individuals in the ACCEPT and PHOENIX trials, respectively.}
\label{tab:accept_phoenix_tab1}
\renewcommand{\arraystretch}{1.3}
\resizebox{\columnwidth}{!}{%
\begin{tabular}{l|cccc|cccc}
\hline
\multicolumn{1}{c|}{\textbf{Trial}} & \multicolumn{4}{c|}{\textbf{ACCEPT}} & \multicolumn{4}{c}{\textbf{PHOENIX}} \\ \hline
\multicolumn{1}{c|}{\textbf{Covariates}} & \textbf{\begin{tabular}[c]{@{}c@{}}Etanercept\\ $(n = 347)$\end{tabular}} & \textbf{\begin{tabular}[c]{@{}c@{}}Ustekinumab 45mg\\ $(n = 209)$\end{tabular}} & \textbf{\begin{tabular}[c]{@{}c@{}}Ustekinumab 90mg\\ $(n = 347)$\end{tabular}} & \textbf{\begin{tabular}[c]{@{}c@{}}Missing\\ $n$ (\%)\end{tabular}} & \textbf{\begin{tabular}[c]{@{}c@{}}Placebo\\ $(n = 255)$\end{tabular}} & \textbf{\begin{tabular}[c]{@{}c@{}}Ustekinumab 45mg\\ $(n = 255)$\end{tabular}} & \textbf{\begin{tabular}[c]{@{}c@{}}Ustekinumab 90mg\\ $(n = 256)$\end{tabular}} & \textbf{\begin{tabular}[c]{@{}c@{}}Missing\\ $n$ (\%)\end{tabular}} \\ \hline
\textbf{Age, mean (SD)} & 41.20 (13.47) & 41.11 (12.95) & 40.11 (12.83) & 1 (0.1) & 44.83 (11.32) & 44.85 (12.48) & 46.20 (11.27) & 0 (0.0) \\
\textbf{Male, $n\;(\%)$} & 246 (70.9) & 133 (63.6) & 233 (67.3) & 1 (0.1) & 183 (71.8) & 175 (68.6) & 173 (67.6) & 0 (0.0) \\
\textbf{BSA affected, mean (SD)} & 23.77 (13.91) & 26.72 (17.79) & 26.08 (17.63) & 1 (0.1) & 27.67 (17.43) & 27.18 (17.49) & 25.18 (15.03) & 0 (0.0) \\
\textbf{PASI, mean (SD)} & 18.64 (6.15) & 20.49 (9.18) & 19.87 (8.38) & 1 (0.1) & 20.41 (8.60) & 20.50 (8.60) & 19.75 (7.64) & 0 (0.0) \\
\textbf{Psoriatic arthritis, $n\;(\%)$} & 95 (27.4) & 62 (29.7) & 95 (27.4) & 0 (0.0) & 90 (35.3) & 74 (29.0) & 95 (37.1) & 0 (0.0) \\
\textbf{PGA, $n \; (\%)$} &  &  &  & 1 (0.1) &  &  &  & 1 (0.1) \\
\multicolumn{1}{r|}{\textbf{Mild}} & 0 (0) & 0 (0) & 1 (0.3) &  & 12 (4.7) & 17 (6.7) & 13 (5.1) &  \\
\multicolumn{1}{r|}{\textbf{Moderate}} & 199 (57.3) & 111 (53.1) & 201 (58.1) &  & 131 (51.4) & 124 (48.6) & 133 (52.2) &  \\
\multicolumn{1}{r|}{\textbf{Marked}} & 135 (38.9) & 87 (41.6) & 135 (39.0) &  & 102 (40.0) & 100 (39.2) & 96 (37.6) &  \\
\multicolumn{1}{r|}{\textbf{Severe}} & 13 (3.7) & 11 (5.3) & 9 (2.6) &  & 10 (3.9) & 14 (5.5) & 13 (5.1) &  \\
\textbf{Ever-smoker, $n\;(\%) $} & 242 (69.7) & 129 (61.7) & 235 (67.7) & 0 (0.0) & 143 (56.1) & 159 (62.4) & 150 (58.6) & 0 (0.0) \\
\textbf{Diabetes mellitus, $n\;(\%) $} & 38 (11.0) & 16 (7.7) & 30 (8.6) & 0 (0.0) & 38 (14.9) & 28 (11.0) & 26 (10.2) & 0 (0.0) \\
\textbf{Previous interventions, $n\;(\%)$} &  &  &  &  &  &  &  &  \\
\multicolumn{1}{r|}{\textbf{Phototherapy}} & 223 (64.5) & 138 (66.3) & 230 (66.3) & 2 (0.2) & 150 (58.8) & 173 (67.8) & 169 (66.0) & 0 (0.0) \\
\multicolumn{1}{r|}{\textbf{Systemic agents$^\dagger$}} & 208 (60.1) & 131 (63.3) & 191 (55.2) & 4 (0.4) & 142 (55.7) & 141 (55.3) & 141 (55.1) & 0 (0.0) \\
\multicolumn{1}{r|}{\textbf{Biologic agents$^\ddagger$}} & 41 (11.8) & 26 (12.5) & 36 (10.4) & 2 (0.2) & 128 (50.2) & 134 (52.5) & 130 (50.8) & 0 (0.0) \\ 
\textbf{Week 12 PASI 75, $n\;(\%)$} & 198 (58.4) & 142 (70.3) & 255 (74.8) & 21 (2.3) & 10 (4.0) & 171 (67.1) & 170 (68.0) & 8 (1.0) \\ \hline
\end{tabular}%
}\vspace{4pt}
\begin{flushleft}
\footnotesize
\noindent $^\dagger$Systemic agents included psoralen plus ultraviolet A therapy, methotrexate, acitretin, and cyclosporin.\\
\noindent $^\ddagger$Biologic agents in the ACCEPT trial included alefacept, efalizumab, adalimumab, and infliximab. Biologic agents in the PHOENIX trial included the same agents as in ACCEPT plus etanercept. \end{flushleft}
\end{table}
\end{landscape}


\begin{table}[htbp]
\centering
\caption{Causal mean differences for achieving PASI 75 comparing treatment assignment to etanercept 50mg versus placebo, within the population underlying the index ACCEPT trial. $\widehat \psi$ is an estimate for $\psi$, the observed data functional under transportability in mean. $\widehat \phi_1$ and $\widehat \phi_2$ are estimates for $\phi_1$ and $\phi_2$, the observed data functionals obtained under transportability in effect measure, using ustekinumab 45mg and ustekinumab 90mg as their respective common comparators.}
\label{tab:accept_phoenix_causalcontrast}
\renewcommand{\arraystretch}{1.3}
\begin{tabular}{@{\hskip 1pt}l@{\hskip 8pt}c@{\hskip 8pt}c@{\hskip 8pt}c@{\hskip 1pt}}
\toprule
\textbf{Estimator} & $\widehat \psi \; (95\% \;\text{CI})$ & $\widehat \phi_1 \; (95\% \;\text{CI})$ & $\widehat \phi_2 \; (95\% \;\text{CI})$ \\
\midrule
Outcome modeling (OM)        & 0.51 (0.42, 0.58) & 0.55 (0.41, 0.65) & 0.50 (0.36, 0.60) \\
Weighting 1 (W1)             & 0.53 (0.46, 0.59) & 0.54 (0.40, 0.69) & 0.51 (0.36, 0.66) \\
Weighting 2 (W2)             & 0.53 (0.46, 0.59) & 0.56 (0.45, 0.67) & 0.57 (0.46, 0.67) \\
Augmented weighting 1 (AW1)  & 0.52 (0.42, 0.58) & 0.56 (0.42, 0.66) & 0.49 (0.35, 0.60) \\
Augmented weighting 2 (AW2)  & 0.52 (0.42, 0.58) & 0.56 (0.42, 0.66) & 0.49 (0.35, 0.60) \\
Augmented weighting 3 (AW3)   & 0.51 (0.41, 0.58) & 0.55 (0.41, 0.66) & 0.50 (0.36, 0.60) \\
\bottomrule
\end{tabular}
\end{table}

\newpage
\section*{ACKNOWLEDGMENTS} 
This study, carried out under YODA Project 2024-0356, used data obtained from the Yale University Open Data Access Project, which has an agreement with JANSSEN RESEARCH \& DEVELOPMENT, L.L.C.. The interpretation and reporting of research using this data are solely the responsibility of the authors and does not necessarily represent the official views of the Yale University Open Data Access Project or JANSSEN RESEARCH \& DEVELOPMENT, L.L.C.. Approval for the conduct of the clinical study was obtained from the Harvard Longwood Medical Area Institutional Review Board, and data use subject to a data use agreement. We thank Brandon Spiegel, a former student at the Harvard T. H. Chan School of Public Health, for his early technical input regarding the simulation study.

\clearpage

\section*{REFERENCES}
\printbibliography[heading=none]

\end{document}